# GUST: Graph Edge-Coloring Utilization for Accelerating Sparse Matrix Vector Multiplication

Armin Gerami
agerami@umd.edu
Perceptual Interfaces and Reality Laboratory
University of Maryland, College Park
College Park, MD, USA

Bahar Asgari
bahar@umd.edu
University of Maryland, College Park
College Park, MD, USA

**Abstract**

Sparse matrix-vector multiplication (SpMV) plays a vital role in various scientific and engineering fields, from scientific computing to machine learning. Traditional general-purpose processors often fall short of their peak performance with sparse data, leading to the development of domain-specific architectures to enhance SpMV. Yet, these specialized approaches, whether tailored explicitly for SpMV or adapted from matrix-matrix multiplication accelerators, still face challenges in fully utilizing hardware resources as a result of sparsity. To tackle this problem, we introduce GUST, a hardware/software co-design, the key insight of which lies in separating multipliers and adders in the hardware, thereby enabling resource sharing across multiple rows and columns, leading to efficient hardware utilization and ameliorating negative performance impacts from sparsity. Resource sharing, however, can lead to collisions, a problem we address through a specially devised edge-coloring scheduling algorithm. Our comparisons with various prior domain specific architectures using real-world datasets shows the effectiveness of GUST, with an average hardware utilization of 33.67%. We further evaluate GUST by comparing SpMV execution time and energy consumption of length-256 and -87 GUST with length-256 1-dimensional systolic array (1D), achieving an average speedup of 411× and 108×, and energy efficiency improvement of 137× and 148×, respectively. To asses the implementation aspect, we compare resource consumption of GUST with 1D as a baseline through FPGA synthesis. Length-256 GUST uses the same number of arithmetic units as length-256 1D, while length-87 GUST uses considerably less. We also compare GUST with Serpens, a state-of-the-art FPGA-based SpMV accelerator, with GUST achieving lower execution time on seven out of nine matrices and lower energy consumption on four.

## 1 Introduction

Data in the form of sparse matrices is ubiquitous in many scientific and engineering applications, such as electronic structure simulation, graph analysis, computational fluid dynamics, recommendation models, and machine learning. Most of the problems in these fields, from solving the Schrödinger equation to calculating the kernel function of a support vector machine, can be solved through sparse matrix-vector multiplication (SpMV). However, without employing techniques such as sparsity-aware hardware or preprocessing sparse data to make it hardware-compatible, zero elements would consume the same amount of processing time as non-zero (NZ) elements, resulting in poor hardware utilization and high execution time. To clarify, we define hardware utilization as the ratio of average number of arithmetic units performing NZ operations in each cycle to total number of arithmetic units.

Prior work have followed approaches such as taking advantage of the high data-reuse rate of 1-dimensional (1D) [22] and 2-dimensional systolic arrays (2D) [10, 11, 33], efficient handling of sparse data structures of balanced adder trees (AT) [1, 3, 9, 25, 27], mapping the hardware to a custom calculational kernel [8, 23, 31], or employing a hardware/software co-design that involves software-based techniques such as preprocessing, prefetching, and novel storage formats working in tandem with custom hardware design [2, 8, 12, 16, 21, 29–32, 36] to accelerate sparse matrix multiplication. In simple terms, all the mentioned work deploy various mechanism to reduce the number of zero elements in the input stream. Although they successfully do so and improve their desired performance metric, their hardware utilization is still far from ideal leaving room for improvement.

In this paper, we introduce GUST, a hardware/software co-design SpMV accelerator developed with the purpose of maximizing hardware utilization. Specifically, given a number of arithmetic units, we aim to minimize execution time. To achieve this goal, we design GUST on the principle that sharing hardware resources among multiple rows and columns would lead to reducing the stalls introduced by irregular sparsity patterns, increasing hardware utilization. The hardware of GUST consists of a number of multipliers connected to a number of adders through a crossbar connector, which enables resource sharing through separating partial product and partial sum calculations. Resource sharing, however, makes SpMV prone to collisions. To prevent collisions, we propose a scheduling based on bipartite-graph Edge-Coloring, an example of using combinatorics to solve computer architecture problems. For demonstration purposes, we adapt and evaluate GUST with an FPGA as the target hardware, and provide comparison with other FPGA-based designs. Similar to prior FPGA-based SpMV accelerators [1, 10–12, 29–31] we perform preprocessing on the matrix, which is our scheduling step.

To justify the rationale behind our design, we first introduce well-known architectures for matrix multiplication, systolic arrays and adder trees, and recent work repurposing them for SpMV in Section 2, and comprehensively explains GUST in Section 3. Then, based on the experimental setup summarized in Section 4, we compare GUST's performance with the mentioned work with a focus on hardware utilization in Section 5. We further evaluate GUST by comparing the end-to-end SpMV wall-clock time and energy consumption against Serpens [29], a state-of-the-art FPGA-based SpMV accelerator. For our evaluation setup, we use a combination of synthetic and real-world matrices, covering a wide range of matrix densities, dimensions and structures, and use `Alveo U280` FPGA for synthesis, which we explain in more detail in Section 4. Section 6 briefly mentions some of the relevant work, and the paper concludes with Section 7. In summary, we achieve the following results:

- On real-world sparse matrices, length-256 GUST achieves average hardware utilization of 33.67%.
- Taking length-256 1D systolic arrays as baseline, length-256 GUST achieves a speedup and energy efficiency gain of 411× and 137×, respectively. Moreover, length-87 GUST has a considerably less hardware and energy consumption than length-256 1D, while achieving average speedup of 108× and energy efficiency gain of 148×.
- Compared to Serpens [29], among the nine real-world matrices we evaluated, GUST achieved lower execution time for seven, and lower energy for four of the matrices.

## 2 Previous Works & Challenges

This section discusses four approaches for performing SpMV that fall under the category of systolic arrays and adder trees: Flex-TPU [10], baseline 1D [17], baseline AT [4], and Fafnir [1]. We analyze their shortcomings to uncover some of the challenges governing efficient SpMV. Table 1 shows the qualities of these designs as well as GUST.

### 2.1 Systolic Arrays

First introduced in 1978 [17], systolic arrays consist of a grid of processing elements (PEs) that execute operations on data in a highly parallelized and pipelined fashion. The PEs are linked to one another, creating a flow of data as it moves from one PE to the next. The purpose of systolic arrays is to leverage spatial locality in matrix multiplications. Spatial locality refers to the tendency for utilized data to be located in close proximity in memory. As a result, data is forwarded between PEs without accessing memory multiple times.

*2D Systolic Arrays*– Given 2D systolic arrays are available for the purpose of matrix-matrix multiplication [14, 15], recent studies, Sparse TPU [11] and Flex-TPU [10], suggest repurposing a 2D TPU to accelerate SpMV while reducing memory access. In the case of Flex-TPU [10], the most recent 2D used for SpMV, each PE is either a Normal PE or a Separator. Unlike baseline 2D, only the NZ matrix values are mapped to PEs, and they are stored in the Normal PEs. Since we only store NZ elements, matrix elements from different rows may be mapped in the same row of the systolic array; therefore, we use Separator PEs to keep track of the matrix rows. Calculating SpMV, as shown in Figure 1(a), involves a left-to-right loading of matrix elements and separator signals during the reconfiguration process. These separator signals identify Processing Elements (PEs) that function as separators. Then, the input vector elements are streamed top-to-bottom and each Normal PE checks the index of the vector element against the column index of its stored matrix element. If there is a match, it performs multiplication and forwards the result to its right neighbor. The Separator PEs accumulate the partial results received from their left-side neighbor and store the result (calculation process). Finally, the dump signal enters through the top-left PE and flows through the PEs. The Separator PEs dump their stored value once they receive the dump signal (dump process). If the matrix size, $m \times n$ is bigger than those of Flex-TPU, $l \times l$, the SpMV is calculated through partitioning, where each partition requires approximately $3 \times l$ cycles ($l$ for reconfiguration, $l$ for calculation, and $l$ for dump) to complete. Given the overhead incurred by PEs that are idle for non-matching indices, and those of reconfiguration and dump processes, Flex-TPU is of low hardware utilization.

*1D Systolic Array*– As shown in Figure 1(b), to calculate matrix-vector multiplication, the elements of the input matrix and the input vector enter the PEs top-to-bottom and left-to-right, respectively. At each cycle, each PE multiplies its entries, accumulates the result to its stored value, stores the result of the sum, forwards its left-side entry to its right-side PE, and dumps its stored value upon receiving the dump

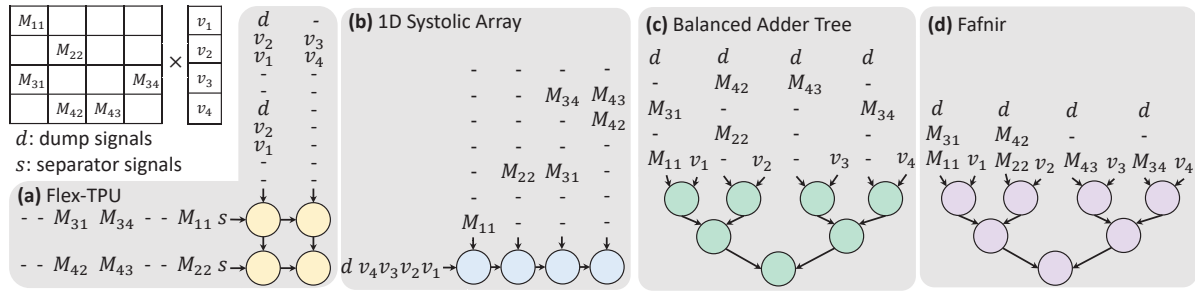

**Figure 1.** Mechanism of SpMV in Prior Work– **(a)** Flex-TPU: At each iteration, first the matrix values and Separators ($s$) are loaded, then the vector values flow through the array, and then the dump signal ($d$); **(b)** 1D; **(c)** AT; **(d)** Fafnir: Each column of matrix enters through one leaf.

**Table 1.** Qualities of related work and GUST for calculating an SpMV for an $m \times n$ matrix. Each design has a length of $l$.

| Design<br>Criteria | Flex-TPU [10] | 1D Systolic Array [17] | Balanced Adder Tree [4] | Fafnir [1] | GUST (our work) |
|---|---|---|---|---|---|
| Hardware | grid of $l \times l$ PEs (2D systolic array) | strip of $l$ PEs | binary tree with $l$ multiplier leaves and $l-1$ reduction nodes | binary tree with $l$ multiplier leaves and $l-1$ reduction nodes | $l$ multipliers and $l$ adders connected by a reconfigurable crossbar connector |
| Execution Time (#cycles) | $\sim 3 \frac{\text{\#NZ elem.}}{l}$ | $\frac{m \times n}{l} + l + 1$ | $\frac{m \times n}{l} + log(l) + 1$ | at least #NZ elem.$\times \frac{log(l)}{4}$ | $3 \times \frac{\text{\#NZ elem.}}{l}$ * |
| Hardware Utilization | 1.45%* | 0.08%* | 0.08%* | 4.67%* | 33.67%* |

*The values reported are based on the geometric average of empirical results for real-world matrices.

signal. Given that each matrix element requires one cycle of processing on a PE and that the majority of the input stream for SpMV are zeros, 1D has poor hardware utilization.

### 2.2 Adder Trees

***Balanced Adder Tree*–** An AT of length-$l$ has $l$ multipliers in the first layer, $l/2$ adders in the second, $l/4$ in the third, and so on (a total of $l-1$ adders) [4]. A pair of multipliers or adders in each layer are connected to an adder in the layer below. As illustrated in Figure 1(c) each multiplier receives two inputs, multiplies them and forwards the result to the adder it is connected to. Similarly, each adder accumulates the two inputs and forwards the result to the layer below. To calculate matrix-vector multiplication, at each iteration, we map a row of the input matrix and the input vector to the inputs of the multipliers and calculate the dot product. Since each element occupies a multiplier and adder for one cycle, and most input elements are zeros, AT demonstrates poor hardware utilization in SpMV.

***Fafnir*–** Fafnir [1] is a tree-like structure of PEs that uses LIL format and can be used to accelerate sparse gathering and SpMV. A Fafnir with length-$l$ would be a binary tree with $l$ leaf nodes, and a depth of $log(l)$, totaling in $2l-1$ PEs. To calculate SpMV, as shown in Figure 1(d), each leaf node receives an input matrix element, the corresponding row index, and the input vector element it should be multiplied with. The leaf node multiplies matrix and vector elements, forwarding the partial product and row index to its parent. Each non-leaf node receives partial products and indices potentially from different rows. It checks for index matches, accumulating and reducing values accordingly - only forwarding the accumulated result instead of all matching values. The node then forwards the reduced results, unreduced inputs, and corresponding indices. In the worst case, the root node performs $l/2$ accumulations per cycle. To prevent stalls, it has $l/2$ adders. Similarly, the $(log(l)-1)$-th layer nodes each have $l/4$ adders, $(log(l)-2)$-th layer $l/8$, and so on. This makes the maximum attainable hardware utilization of Fafnir to be $4/log(l)$, but in practice, it is significantly lower.

### 2.3 Challenges

We explained why each design achieves low hardware utilization, and we confirm this empirically in Section 5. A key factor contributing to these limitations lies in the conventional static resource allocation. For example, at each iteration in 1D, each PE is allocated to a matrix row, and cannot be utilized by another row, even if it is idle. Similarly in Flex-TPU, each PE is allocated to an element. In AT/Fafnir, each multiplier is allocated to a single row/column at each iteration, and each layer 2 adder allocated to 2 rows/columns, layer 3 to 4, and so on. Having a predefined resource allocation limits our ability to utilize the hardware based on the sparsity of each row/column.

## 3 GUST

GUST is a hardware/software co-design with the main goal of achieving a high hardware utilization for SpMV. The hardware of GUST facilitates this by enabling resource sharing,

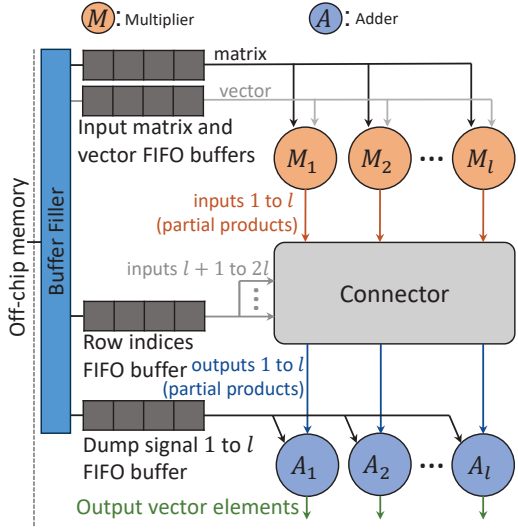

**Figure 2. Overview of GUST Architecture**– The Buffer Filler retrieves the matrix in the scheduled format from off-chip memory, and fills four sets of buffers: the matrix and vector elements, the row indices, and the dump signals.

while the software schedules, load balances, and maps the SpMV inputs to the hardware inputs.

### 3.1 Key Insight

High hardware utilization is associated with a dense input stream; the denser the input stream the less stalls/zero operations. Therefore, our goal is to reshape the input matrix into a dense stream. Doing so will result in changing the order of the matrix elements. As a result, we need to make sure that each matrix element is multiplied by the correct vector element, and that the resulting partial product is accumulated with the correct partial sum. One way to do so is to separate the multipliers and adders. We then shape the multipliers' input stream in a way so that correct values get multiplied together, and use a crossbar connector between the multipliers and adders to control the adders' input stream, so that correct partial products get accumulated with correct partial sums. Thus the hardware utilization is increased through having a dense input stream, and arithmetic units are shared among multiple rows and columns. As explained with detail in Section 3. 3, resource sharing is prone to collisions. In order to eliminate collisions, we propose our bipartite-graph Edge-Coloring scheduler. We further improve the performance of GUST with a simple sort-based load balancing.

### 3.2 Hardware & Core Mechanisms

**Hardware**– As illustrated in Figure 2, the hardware of GUST consists of three layers. A length $l$ GUST has $l$ multipliers, $l$ adders, and a crossbar connector:

-Multipliers: Each multiplier has two inputs; one dedicated to the input matrix elements, and one to the input vector. In each cycle, the multipliers multiply their inputs and forward the partial products to the connector. The first input of the $j$-th multiplier is an element from the $j$-th column of the input matrix, the second input the $j$-th element of the input vector, and the output the $j$-th input of the connector.

-Crossbar Connector: The connector has two sets of $l$ inputs, and one set of $l$ outputs. The first $l$ inputs are the partial products that are propagated from the multipliers. The outputs of the connector are also partial products, connected to adders. The second $l$ inputs are indices that range from 1 to $l$, and regulate which partial product entry is forwarded to which adder. Specifically, the $j$-th entry (a partial product) is forwarded to the value of the $(j + l)$-th input (an index).

-Adders: Each adder has two inputs; one forwarded from connector (a partial product), and one dedicated to the dump signal. In each cycle, the adders accumulate the input from the connector with their stored value, and store the result. The input buffers are populated with the help of the Buffer Filler element, which utilizes an on-chip memory. The process works as follows: First, the vector is forwarded from off-chip memory to the Buffer Filler. Then, the buffers are filled in a two-step pipelined fashion. Step 1: A partition of the scheduled matrix is forwarded from off-chip memory to the Buffer Filler's on-chip memory. Step 2: The Buffer Filler fills the input buffers using the matrix and vector elements stored in its on-chip memory. The Buffer Filler's existence is necessary in order to facilitate this pipelined process.

**Data Flow**– To calculate an element of the output vector, $y_i = \Sigma_j M_{ij} v_j$, each partial product, $M_{ij} v_j$, is calculated as follows. $M_{ij}$ and $v_j$ enter the $j$-th multiplier. The multiplier calculates $M_{ij} v_j$ and forwards it to the $j$-th input of the connector. The row index value, $i$, enters the connector as the $(j+l)$-th input. Based on the index value, $i$, the connector forwards $M_{ij} v_j$ to the $i$-th adder. The adder accumulates the partial product to its stored value. The dumped value by the $i$-th adder will be $y_i = \Sigma_{j, M_{ij} \neq 0} M_{ij} v_j$. Each of the four inputs (matrix and vector elements, row indices, and dump signal) are connected through an individual FIFO buffer.

If the dimension of the input matrix, $m \times n$, is bigger than the length of GUST, $l$, SpMV is done through windowing, meaning at each iteration, a set of $l$ rows enter the multipliers, and once all of the NZ elements in the $l$ rows have gone through the accelerator, the adders dump their values, and the next $l$ rows enter the multipliers (i.e., matrix enters set-of-rows by set-of-rows). In addition, since $n$, number of columns, is bigger than $l$, matrix elements enter the multipliers with respect to their column index $mod\ l$, where $mod$ is the modulo operator. That is, the matrix elements in the $j$-th, $(j + l)$-th, $(j + 2l)$-th, ..., $(j + (\frac{n}{l} - 1)l)$-th columns enter through the $j$-th multiplier, where $1 \leq j \leq l$ (we refer to segments of the $j$-th, $(j + l)$-th, ... columns within each window as column segments). Through windowing, resources are shared across both multiple rows and multiple columns. Figure 3 provides a step-by-step example for a length-4 GUST.

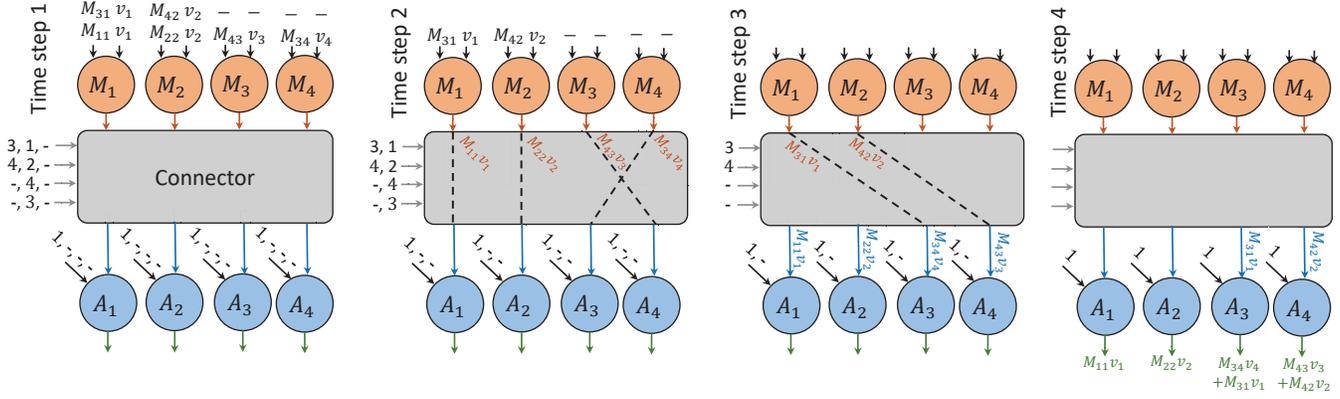

**Figure 3. Hardware Mechanism of GUST**– Time step 1 shows how the matrix and vector elements from the example in Figure 1 enter the hardware. The dashed-lines in time steps 2 and 3 display which input of connector is forwarded to which output. By receiving dump signal in timestep 4, adders dump their stored value.

### 3.3 Scheduling and Load Balancing

The example in Figure 3 was handpicked in the sense that in no cycle were there any two matrix elements from the same row entering the multipliers, or equivalently, no two connector inputs were destined for the same connector output. In general, this is not the case and multiple matrix elements from the same row may enter in the same cycle and cause a collision in the outputs of the connector, losing the partial products. The naive method of preventing collisions is to simply not forward the values from the buffers whenever there is a collision and add a stall. However, this method introduces a high number of stalls, resulting in significant inefficiency. In fact, empirical results demonstrate that for $16384 \times 16384$ matrices with uniform distribution, GUST using naive scheduling has a performance worse than 1D for densities exceeding 0.008. As a remedy, we propose our scheduling algorithm, which is based on the notion of Edge-Coloring a bipartite-graph.

*Edge-Coloring a bipartite-graph*– First, let us showcase a simple problem to see how Edge-Coloring could be of any avail. As illustrated in Figure 4, assume we have an $n \times m$ table, and its cells can either be empty or take a value. We want to fill the table with the set of values $V = S_1, S_2, ..., S_n$, where $S_k = v_{k,i}, v_{k,i+1}, ..., v_{k,j}$, and $0 \leq i \leq j \leq m$. The values in each $S_k$ are unique within $S_k$ but not necessarily within $V$. We want to fill the $k$-th column of the table with $S_k$ in a way that each value of a cell would be unique in its row. To do so, assume we have a bipartite-graph where the left-side vertices represent the columns of the table, the right-side vertices the values, and the edge colors the rows. Therefore, if an edge connects the $j$-th left-side vertex to the $v$-th right-side vertex and with the color of $i$, the cell in the $i$-th row and $j$-th column has the value of $v$. By coloring the edges in a way that no two adjacent edges would have the same color, we fill the table and satisfy the mentioned conditions. According to Vizing's theorem [34], the minimum number of colors needed is the maximum degree of the graph, which in our case would be $max\{max_k\{|S_k|\}, r\}$, where $r$ is the the maximum occurrences of a value in $V$.

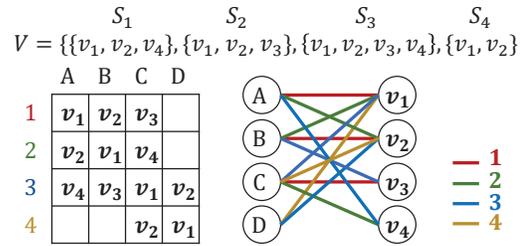

**Figure 4. Edge Coloring**– Filling the table such that $S_j$ fills the $j$-th column and no repeated values exist in any row. Since each vertex has at most one color connected to it, there are no repeating values in any given column or row.

*Applying Edge-Coloring a Bipartite-Graph to GUST*– The collision issue we are facing is analogous to the showcased problem, and we apply a similar approach to address it. Here, matrix elements enter multipliers based on their column index, and are directed towards adders based on their row index. Since the column and row indices of the matrix elements are predetermined and out of our control, we represent them as the right- and left-side vertices of a bipartite-graph. An edge between the $i$-th left-side vertex and the $j$-th right-side vertex represents the matrix element from row $i$ and column $j$. Since the color of edges can be deliberately assigned, we use colors as our scheduling metric. Only one matrix element can enter a multiplier at each cycle, therefore, the scheduling must be in a manner that no two elements from the same column end up with the same time slot, meaning no two edges connected to a right-side vertex can have the same color. Similarly, no two elements from the same row can enter the adder at the same time slot, meaning no two edges connected to a left-side vertex can have the same color. Our problem then becomes, given a bipartite-graph, where the left-side vertices represent the matrix rows, the right-side vertices the matrix columns, and the edge colors the scheduled position in the input buffer, find a coloring such that no vertex has two edges with the same color connected to it. This ensures that no two elements from the same row would occupy the same position in the input buffers, eliminating all collisions. In terms of relevance to the hardware, since each adder is responsible for accumulating the partial products from a single row, and that each column enters through a single column, each left-side vertex corresponds to an adder, and right-side vertex to a multiplier. For example, the matrix element $M_{ij}$ will be represented by an edge connecting the $i$-th left-side vertex to the $j$-th right-side vertex of the bipartite-graph, the edge color will be its position in the input buffer of the $j$-th multiplier, and it will be directed to the $i$-th adder.

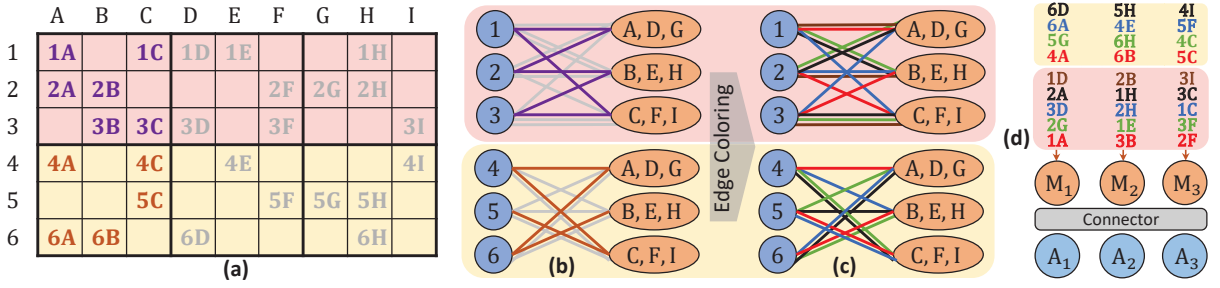

**Figure 5. Edge Coloring in GUST–** **(a)** A sparse matrix operand for an SpMV. **(b)** The bipartite-graph representation of the sparse matrix, in which left/right nodes correspond to rows/columns (adders/multipliers). The top graph (red background) represents the first three rows of the matrix, and the bottom one (yellow background) the last three. Some of the edges have been colored to illustrate their corresponding matrix element for clarification. **(c)** The Edge-Color-scheduled graphs where the edge color corresponds to the position of the matrix element in the FIFO buffer. No two edges with the same color are connected to the same vertex. **(d)** The scheduled matrix elements entering the multipliers of GUST.

If the dimension of the input matrix, $m \times n$, is bigger than the length of GUST, $l$, the scheduling is done through windowing; meaning first, the first $l$ rows are scheduled, then the second $l$ rows and so on. In addition, since $n$ is bigger than $l$, the scheduling will be done with respect to the column $mod$ $l$ of the matrix elements; meaning multiple matrix columns will be represented by the same right-side vertex. To be specific, the $j$-th, $(j+l)$-th, $(j+2l)$-th, ..., $(j+(\frac{n}{l}-1)l)$-th columns are represented by the $j$-th right-side vertex, where $1 \leq j \leq l$. This also means that it is possible to have multiple edges connecting the same pair of vertices.

Figure 5 provides an example of scheduling for an SpMV with a 6×9 matrix using a length-3 GUST. The top right-side vertex in Figure 5(b) and (c) represent the first, fourth, and seventh columns (A, D, G), the middle vertex the second, fifth, and eighth columns (B, E, H), and the bottom vertex the third, sixth, and ninth columns (C, F, I). As a result, we observe that elements from the first, fourth, and seventh columns are directed to the first multiplier, the second, fifth, and eighth columns to the second multiplier, and the third, sixth, and ninth columns to the third multiplier.

**GUST Scheduling Algorithm–** The scheduling for a length-$l$ GUST generates three outputs: the matrix in scheduled format $M_{sch}$, and the $Row_{sch}$ and $Col_{sch}$ matrices. $M_{sch}$ is an $l$ by $C_{total}$ matrix, where $C_{total}$ is the total number of colors needed. $M_{sch}$ contains elements of the original matrix in a rearranged and compressed manner. $Row_{sch}$ and $Col_{sch}$ are also $l$ by $C_{total}$ matrices. They contain the original row $mod$ $l$ and column indices of the elements in $M_{sch}$, and are used to control the data-flow in the FPGA during SpMV. These matrices can be viewed as a compressed storage format similar to the Coordinate format.

```
1 //l = Gust len, E = edge set, a 2-d variable
2 clr = 0;
3 While(E not Empty)
4   matching = []
5   //iterate through the left-side vertices
6   for(i = 1 to l)
7     //iterate through the edges connecetd
8     for(k = 1 to E[i].size())
9       if(E[i][k] mod l not in matching)
10        color[i][E[i][k]] = clr
11        matching.insert(E[i][k] mod l)
12        delete E[i][k]
13        break
14  clr += 1
```

**Listing 1.** Edge-Coloring Algorithm

To fill these three matrices, we first need to find the scheduling. For an $m \times n$ matrix and a GUST of length $l$, the matrix is represented by $m/l$ sets of edges, and we schedule each set independently. Each set connects $l$ left-side vertices to $l$ right-side vertices, with the possibility of multiple edges connecting the same two vertices. To do so, we start with $clr = 0$. At each iteration, we find a matching (a matching is a subset of edges which share no vertex), assign the color value of the matching as the current $clr$ value, remove the matching from the set, and update color value as $clr = clr + 1$. We repeat these steps until all of the edges have a color assigned to them. Listing 1 is the pseudo-code of our Edge-Coloring algorithm for a set of $l$ rows, where **E** represents a set of edges from $l$ rows of the input matrix. To be specific, **E[i][k] = j** indicates an edge connecting the $i$-th left-side vertex to the *j mod l*-th right-side vertex. This representation allows for a faster Edge-Coloring since when we choose an edge from a left-side vertex to be in the matching, we do not need to consider any edges connected to that vertex (the **break** statement at line 13). Moreover, **color[i][j]** represents the color of the matrix elements in the *i*-th row and *j*-th column.

With the schedule (color of edges) determined, we can now proceed to fill $M_{sch}$, $Row_{sch}$ and $Col_{sch}$. $M_{sch}[i][j]$ contains the matrix element that should enter the *j*-th multiplier at the *i*-th timestep. $Col_{sch}[i][j]$ and $row_{sch}[i][j]$ contain the original column index and row *mod l* index of this element. Using them we keep track of which vector element $M_{sch}[i][j]$ should be multiplied with, and to which adder the result of this multiplication should be forwarded; for instance, $M_{sch}[i][j]$ will be multiplied with the $Col_{sch}[i][j]$-th element of the vector, and the crossbar will connect the *j*-th multiplier to the $row_{sch}[i][j]$-th adder. Listing 2 shows how to fill these matrices with respect to color, where M_sch, row_sch and col_sch refer to $M_{sch}$, $Row_{sch}$ and $Col_{sch}$. To elaborate, if the color of a matrix element in the *i*-th row and *j*-th column is *c*, it will be filled in M_sch[c][j mod l], since it should enter the *j mod l*-th multiplier at the *c*-th time step. Moreover, row_sch[c][j mod l] and col_sch[c][j mod l] will be filled with *i mod l* and *j*. Note that the scheduling for each matrix only needs to be computed once. We can then reuse this schedule to fill the buffers for any SpMV with that matrix, even if the input vector changes. Furthermore, if the matrix changes but the location of NZs remain the same (as it is the case with Jacobian and Hessain matrices), the scheduling (Listing 1) does not need to be repeated, rather $M_{sch}$ (Listing 2) needs to be updated.

```
1 //l = Gust len, m = matrix height
2 // M = input matrix
3 for(i = 1 to m)
4   for(k = 1 to colors[i].size())
5     M_sch[color[i][j]][j mod l]   =   M[i][j]
6     row_sch[color[i][j]][j mod l] =   i mod l
7     col_sch[color[i][j]][j mod l] =   j
```

**Listing 2.** Filling the Output Matrices

***Streaming the Inputs*–** Given $M_{sch}$, $Row_{sch}$ and $Col_{sch}$ matrices, we will now go over how to fill the input buffers. As a reminder, there are four types of input buffers: **matrix**, **vector**, **row indices** and **dump signal** buffers, as shown in Figure 2. We begin with forwarding the input vector to the Buffer Filler, which utilizes on-chip memory. Then, we fill the buffers in a two-step pipelined fashion. Step 1: A partition of the $M_{sch}$, $Row_{sch}$ and $Col_{sch}$ matrices are forwarded from off-chip memory to the Buffer Filler's on-chip memory. Step 2: The Buffer Filler fills the input buffers as follows: The **matrix** and **row indices** buffers are filled directly with $M_{sch}$ and $Row_{sch}$ respectively. The **vector** buffers are filled using the vector stored in the on-chip memory, and with respect to $Col_{sch}$, ensuring that each matrix element is multiplied with the correct vector element. Specifically, the **vector** buffer of the *j*-th multiplier at timestep *i* is filled with the $Col_{sch}[i][j]$-th element of the vector. For a length-*l* GUST and matrix with width *N*, the $M_{sch}$ elements are 32 bit floating point values, the $Row_{sch}$ are $log(l)$ bit values (since they should index 1 to *l*), and the $Col_{sch}$ are $log(N)$ bit values (since they should index 1 to *N*). Assuming the $Col_{sch}$ elements are 32 bits, and that the operating frequency is *f* Hz, the required BW by a length-*l* GUST becomes $(64l + log(l) + 1)f$ bits/s.

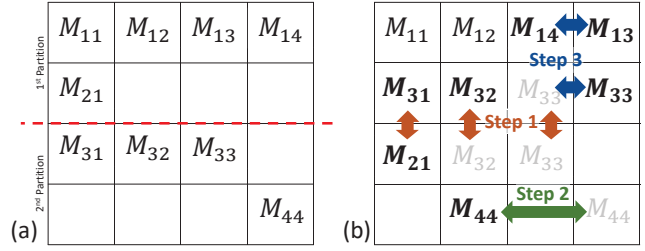

**Figure 6. Load Balancing**– **(a)** Initial matrix. **(b)** The matrix after applying step 1 (sorting the rows, resulting in switching the second and third row), 2 (sorting the column segments, resulting in switching the second and fourth column segments in the second row set), and 3 (rearranging partial columns, resulting in switching the third and forth column segments) of the load balancing algorithm.

### 3.4  Statistical Bound

In this section, we find a statistical bound on the hardware utilization to assess the effectiveness of our scheduling approach. The input buffer length required to process a set of *l* rows is equal to the number of colors required $C$, and according to Vizing's theorem [34], this value is

$$C = max\{max_i\{\#\text{NZ in the }i\text{-th row}\},$$
$$max_j\{\sum_{x=0}^{n/l-1} \#\text{NZ in the }(j+x)\text{-th column}\}\}. \quad (1)$$

In other words, the number of colors needed is the maximum number of edges connected to a vertex in the bipartite graph. The execution time in terms of number of cycles is the sum of the number of colors for all of the edge sets (each edge set is a set of rows) plus 2 (GUST has 3 levels). For example, in Figure 5(c) the first three rows require 5 colors and the last three 4, making the total number of cycles 11.

Using Eq. 1, we will derive a statistical bound on the execution time for sparse matrices with a uniform distribution. This bound will then be used to analyze hardware utilization as a function of the probability *p* of each element being NZ. For a length *l* GUST and an $N \times N$ matrix, each row has *k* NZ elements with probability

$$Pr_{row}(k) = \binom{N}{k}p^k(1-p)^{(N-k)}. \quad (2)$$

Assuming $N > 9(1 − p)/p$ (i.e., an average of at least 10 NZs in each row) we can use the Central Limit Theorem and approximate

$$Pr_{row}(k) \sim \mathcal{N}(Np, Np(1-p)), \tag{3}$$

where $\mathcal{N}$ denotes the normal distribution. Since $p$ is small, we can also assume that the number of NZ in each column segment is independent from rows, making the probability of $k$ NZ elements in a column segment

$$Pr_{col}(k) \sim \mathcal{N}(Np, Np(1-p)) \tag{4}$$

as well. We have $l$ rows and $l$ column segments, and since $C$ is the maximum number of NZs in rows and column segments, the value of $C$ is governed by the maximum of $2l$ independent Gaussian distributions. i.e.

$$C = \max_{i, 1 \leq i \leq 2l} X_i, \quad X_i \sim \mathcal{N}(Np, Np(1-p)). \tag{5}$$

To calculate the expectation of $C$, let us first calculate the expectation of $\mathcal{Y} = \max_{i, 1 \leq i \leq 2l} Y_i$, where $Y_i \sim \mathcal{N}(0, \sigma^2)$. Using Jensen's inequality, we have

$$\forall_t: \quad e^{tE[\mathcal{Y}]} \leq E[e^{t\mathcal{Y}}] = E[\max_i \{e^{tY_i}\}]. \tag{6}$$

Using Union bound, we bound the right-hand side of Eq. 6 with

$$E[\max_i \{e^{tY_i}\}] \leq \Sigma_{i=1}^{2l} E[e^{tY_i}] = 2l e^{t^2\sigma^2/2}. \tag{7}$$

Setting $t = \sqrt{2\log(2l)}/\sigma$ we get

$$E[Y] \leq \sigma\sqrt{2\log(2l)}. \tag{8}$$

For $y_i \sim \mathcal{N}(\mu, \sigma^2)$, the right-hand side of Eq. 8 will be $\mu + \sigma\sqrt{2\log(2l)}$ since we are shifting the values with $\mu$. Replacing $\mathcal{Y}$ with $C$ we get

$$E[C] \leq Np + \sqrt{2Np(1-p)\log(2l)}. \tag{9}$$

The expectation of execution time then becomes

$$E[\text{exe.}] = \frac{N}{l}(Np + \sqrt{2Np(1-p)\log(2l)}) + 2 \tag{10}$$

since execution time is the sum of all the colors for the $N/l$ edges sets. Having the expectation of the execution time, we can find the expectation of hardware utilization as

$$E[\text{hardware utilization}] = \frac{\#NZ/l}{E[\text{exe.}]} = \frac{N^2 p/l}{E[\text{exe.}]} \tag{11}$$
$$= \frac{1}{1 + \sqrt{2(1-p)\log(2l)/Np}}.$$

### 3.5 Load Balancing

Value of $C$ in Eq. 1 depends on the maximum of NZs in column segments and rows, and not the total number of NZs. As a result, performance of GUST can be negatively affected by inconsistencies in the number of NZs (#NZ) between column segments or rows within row sets. Consider the example in Figure 6. According to Eq. 1, case (a) requires 7 cycles (4 cycles for the first partition and 3 for the second) to process while case (b) requires 5 cycles (4 cycles for the first partition and 1 for the second). In general, the smaller the standard deviation of #NZ in rows and column segments within row sets, the smaller the execution time. Our load balancing is a simple yet effective three step sorting procedure: **Step 1:** Sort the matrix rows based on #NZ in each row. **Step 2:** Sort the column segments of each partition based on #NZ in each column segment. **Step 3:** For even column segments, reverse the order of columns (e.g. for a length-2 GUST, if the order of sorted column segments are 1, 2, 3, 4, 5, 6, 7, 8, rearrange them to 1, 2, 4, 3, 5, 6, 8, 7). The first step helps with reducing the standard deviation of #NZ in rows and the second and third step in column segments.

## 4 Experimental Setup

**Dataset.** We evaluate GUST based on synthetic and real-world matrices. For the synthetic data, we include matrices with uniform, power-law and k-regular distribution and a dimension of 16, 384 over a density range of $1e^{-4}$ to $5e^{-2}$. The power-law and k-regular synthetic data were acquired from SNAP [18] matrix generator. For real-world data, we used a collection of matrices from the SuiteSparse [7] and SNAP [18] Matrix Collections, with densities ranging from $1e^{-5}$ to $1e^{-1}$.

**Simulation.** To compare the hardware efficiency of GUST with the designs introduced in Section 2, we assumed all the designs have 256 adders and 256 multipliers, except for Fafnir, which has 448 adders and 128 multipliers. The hardware efficiency of the designs were calculated based on the dataflow of each specific matrix.

We use an `Alveo U280` FPGA to synthesis length-256 GUST, which is HBM2 enabled with a maximum memory bandwidth of 460 GB/s. We use float-32 arithmetic precision and target clock frequency of 96 MHz, which is bounded by GUST's longest logic route connecting the two ends of the crossbar. The minimum BW needed is 224 GB/s, as explained in Section 3. 3. **Streaming the Inputs**. Length-256 1D and Serpens are synthesized using the same FPGA.

The energy efficiency gain is calculated by computing the energy consumption as a results of dynamic power, NZ data movements, reads, writes, and arithmetic operations using the following energy numbers for 32 bits in pJ [5, 6]: 64 and 11.84 for reading from off-chip and on-chip memory read, 64 and 16 for off-chip and on-chip memory write, 10 for floating point accumulation and multiplication, 160 and 0.95 for moving data for 1mm for off-chip and on-chip, with the distance being 5 mm, 1 mm, and 129 mm, between off-chip memory and on-chip elements, on-chip elements in 1D, and average distance between on-chip elements in GUST (we use average since GUST has a crossbar connection). The dynamic power consumption values used for length-256 1D, length-256 GUST and length-87 GUST are 35.3, 56.9 and 16.8 Watts respectively, which were measured from the FPGA synthesis. To account for the energy consumption of the first step of GUST's SpMV calculation (forwarding the vector to Buffer Filler), we add the power consumption of GUST times the duration it takes to forward the values.

`Alveo U280` has 32 physical channels whereas we are interested in implementing a length-256 GUST, which has 18,433 logical inputs ($256 \times 32$ for matrix values, $256 \times 32$ for vector values, $256 \times 8$ for index values, and 1 for the dump signal. This means that the off-chip memory cannot be directly connected to the multipliers. As a workaround, we have introduced the Buffer Filler and a pipelined process for streaming the inputs. To elaborate, at each timestep, the HBM2 streams the inputs to the on-chip memory of the Buffer Filler, and the Buffer Filler appropriately forwards the values to the multiplier, crossbar, and adder input buffers to perform the calculations (double buffering). The required on-chip memory to enable this process is equal to the twice the size of the input values in a timestep, which is 36,866 bits (4.5 KB) in here. Given that `Alveo U280` offers 41 MB of on-chip memory, this leaves ample storage for a vector of dimension up to $10^7$, which is 9 times larger than the biggest matrix encountered in our experiments (as a reminder GUST stores the whole input vector as the first step of the SpMV calculation). We should clarify that that GUST's hardware implementation is not limited to FPGAs. We adapted an FPGA implementation for demonstration purposes, but GUST can be implemented on any hardware platform that can provide a set of multipliers and adders, and a crossbar connector.

## 5 Results

This section provides a comprehensive evaluation of GUST. We first compare the hardware utilization and execution time of GUST with the designs introduced in Section 2. Next, we asses GUST from an implementation perspective. We compare length-256 and length-87 GUST with length-256 1D in terms of speedup, energy efficiency gain, and resource consumption. Then, we compare end-to-end SpMV wall-clock time and energy consumption with Serpens [29]. Then, we evaluate how the density and structure of the input matrix affects the performance of GUST. Finally, we talk about scalability of GUST in terms of its length.

### 5.1 GUST vs Other Designs

Figure 7(a) shows the hardware utilization of GUST with naive scheduling (Naive), GUST with Edge-Coloring (EC), GUST with Edge-Coloring and load balancing (EC/LB), and the designs introduced in Section 2 for SpMV over a range of real-world sparse matrices [7]. The results reveal a significant performance gap between GUST with EC/LB and other designs, evident in both hardware utilization and execution time. Specifically, GUST with EC/LB demonstrates an average hardware utilization of 33.67%, which is notably high when dealing with sparse matrices. Moreover, comparing GUST with Naive and GUST with EC/LB reveals a substantial performance gap as well, emphasizing the critical role of Edge-Coloring in GUST's effectiveness. This gap widens with increasing density as collision-induced stalls add up, to the point that GUST with Naive will have a worst performance than 1D and AT. This underscores the importance of our proposed scheduling algorithm in increasing hardware utilization. Without proper scheduling, the resource sharing enabled by the hardware will not result into a noteworthy performance improvement. Figure 7(b) shows the comparison for execution time. The results exhibit a consistent pattern, demonstrating that high hardware utilization translates into reduced execution time.

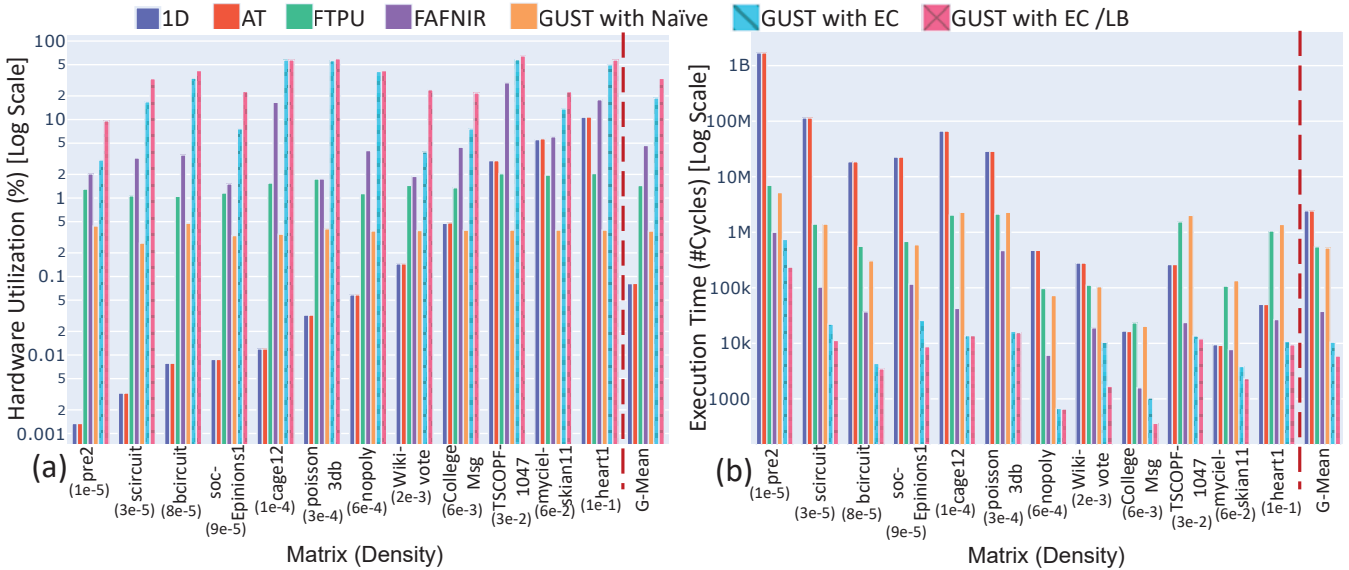

**Figure 7. GUST VS Other Designs** – **(a)** Hardware utilization and **(b)** execution time of various designs for SpMV over real-world matrices with a wide density range of $1e^{-5}$ to $1e^{-1}$. GUST with Edge-Coloring and load balancing (EC/LB).

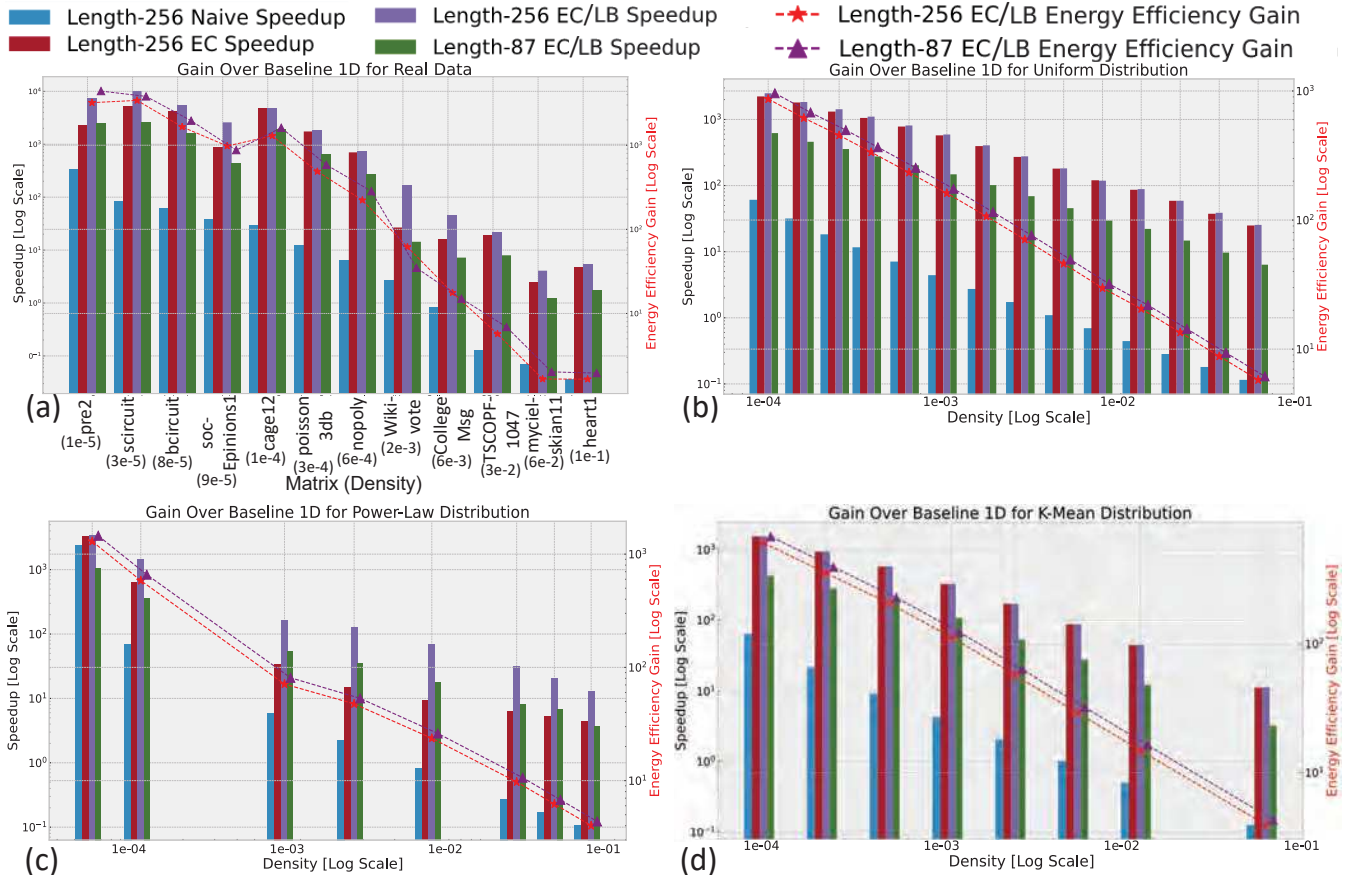

**Figure 8. Performance Gain of GUST Over 1D–** Speedup of length 256 GUST with Naive, Edge-Coloring and Edge-Coloring/Load Balancing, as well as energy efficiency gain of length-256 GUST with Edge-Coloring/Load Balancing over length-256 baseline 1D systolic array on **(a)** real, and synthetic data with **(b)** uniform, **(c)** power-law and **(d)** k-mean distribution.

### 5.2 Performance and Overhead of GUST Over 1D

Figure 8 shows the speedup and energy efficiency gain of length-256 GUST with Naive, GUST with EC, and GUST with EC/LB as well as length-87 GUST with EC/LB over length-256 1D for real-world and synthetic matrices. Table 2 and shows the resource consumption of these designs based on FPGA synthesis. Length-256 GUST with EC/LB uses the same number of arithmetic units as length-256 1D while offering speedup and energy efficiency gain in order of magnitudes. However, it introduces a considerable overhead due to the crossbar connector. One might argue that the main reason for GUST's performance gain over 1D is the mentioned overhead. For this purpose, we also compare length-87 GUST, which uses significantly less hardware resources than length-256

1D while achieving substantial performance gains. Furthermore, looking at the speedup offered by GUST with Naive GUST with and EC/LB we notice a wide gap, once again showing the importance of our scheduling, and that the performance gain does not come from the hardware structure alone. For real-world matrices, length-256 GUST with EC/LB achieves an average speedup of 88× over length-256 GUST with Naive, and 1.8× over length-256 EC. Compared to length-256 1D, length-256 and length-87 GUST with EC/LB achieve speedup of 411× and 108×, and energy efficiency gain of 137× and 148× respectively.

Figure 9 shows the average bandwidth utilization of length-256 and -87 GUST with EC/LB as well as length-256 1D for SpMV over real-world matrices based on the target clock frequency of 96 MHz synthesis. The noticeable difference in bandwidth utilization between GUST and 1D stems from the substantial underutilization of bandwidth by 1D due to sparsity. In fact, the high bandwidth of GUST indicates that we were successful with our goal: to increase hardware utilization through creating a dense input stream.

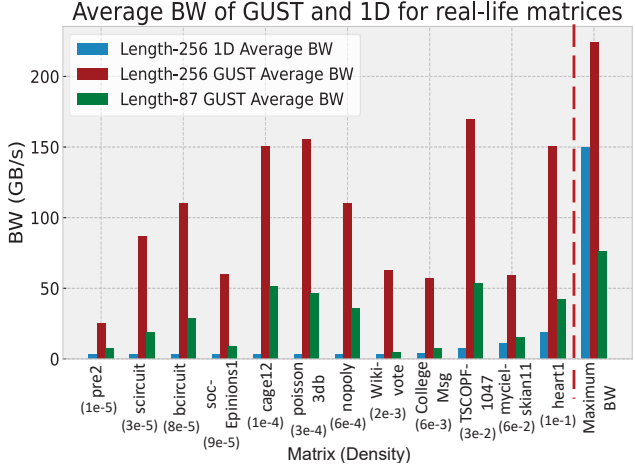

**Figure 9. BW Utilization**– Average BW utilized by length-256 1D and length-256 and -87 GUST with EC/LB. The "Maximum BW" represents maximum possible bandwidth utilization, achievable when all inputs are nonzero.

### 5.3 GUST vs Serpens

We compare GUST with Serpens on nine SuiteSparse [7] and SNAP [18] matrices, and Table 3 shows the details of these matrices. Both GUST and Serpens are synthesized on an `Alveo U280` FPGA with operating frequency of 96 MHz and 223 MHz respectively, and dynamic power of 56.9 and 46.2 Watts. Moreover, both take advantage of preprocessing to reformat the sparse matrix into accelerator-efficient storage. Table 4 shows the execution time, energy consumption and throughput of GUST during the preprocessing and SpMV calculation. As we can see, GUST requires considerably less time and energy to perform its preprocessing. As for the SpMV calculation, even though GUST has a lower operating frequency, it achieves lower execution time for seven out of nine matrices. This is because the scheduling results in a denser matrix format, and since the scheduling is collision-free, it leads to higher FLOPs per cycle. Consequently, the total cycle count and overall wall-clock time is reduced. Furthermore, GUST achieves lower energy consumption for four of the matrices. Despite GUST's higher power consumption, the reduced execution time results in overall lower energy consumption.

**Table 2. Per-Element Resource Consumption of GUST and 1D**– Power and hardware consumption of length-256 1D, length-8, -87 and -256 GUST based on FPGA synthesis.

| Design\Power (Watt) | length-256 1D | length-8 GUST | length-87 GUST | length-256 GUST |
|---|---|---|---|---|
| Static | 3.2 | 2.5 | 3.2 | 3.8 |
| Logic | 3.4 | 0.1 | 1.8 | 14.3 |
| Signals | 2.6 | 0.3 | 3.0 | 8.1 |
| DSP | 0.3 | 0.01 | 0.1 | 0.3 |
| I/O | 25.7 | 0.5 | 8.6 | 30.3 |
| Total | 35.3 | 3.4 | 16.8 | 56.9 |

| Design\Units | length-256 1D | length-8 GUST | length-87 GUST | length-256 GUST |
|---|---|---|---|---|
| Register | 8.2K | 512 | 5.6K | 16.4K |
| Input Buffers | 8.2K | 546 | 6.2K | 18K |
| LUT | 132K | 5K | 5.6K | 888K |
| DSP | 256 | 16 | 174 | 256 |
| I/O Buss | 16K | 802 | 8.9K | 2.7K |
| Maximum BW | 150 GB/s | 5.8 GB/s | 76GB/s | 224GB/s |

**Table 3.** Set of nine real-world matrices with density range of $e^{-5}$ to $e^{-3}$.

| ID | Matrix | Dimension | #NZ | Density |
|---|---|---|---|---|
| (1) | crankseg_2 [7] | 63.8K | 14.1M | $3.4e^{-3}$ |
| (2) | Si41Ge41H72 [7] | 186K | 15.0M | $4.3e^{-4}$ |
| (3) | TSOPF_RS_b2383 [7] | 39.1K | 16.2M | $1.0e^{-2}$ |
| (4) | ML_Laplace [7] | 377K | 27.6M | $1.9e^{-4}$ |
| (5) | mouse_gene [7] | 45.1K | 29M | $1.4e^{-3}$ |
| (6) | coPapersCiteseer [7] | 434K | 21.1M | $1.1e^{-4}$ |
| (7) | PFlow_742 [7] | 743K | 37.1M | $6.7e^{-5}$ |
| (8) | googleplus [18] | 108K | 13.7M | $1.2e^{-3}$ |
| (9) | soc_pokec [18] | 1.63M | 30.6M | $1.2e^{-5}$ |

While the preprocessing times for both GUST and Serpens are orders of magnitude higher than the actual SpMV calculation, this does not pose a significant issue. The reason is that the preprocessing step is a one-time operation for a given sparse matrix, whereas the SpMV calculation needs to be performed repeatedly within iterative solvers. To be specific, these SpMVs usually arise in linear algebra solvers of the form $Ax = y$, where $A$ is the sparse matrix, and the goal is to solve for $x$. Two common approaches exist: reducing the rank of the linear system using techniques like QR decomposition, which requires $O(N^3)$ operations ($N$ being the matrix size), or employing gradient/conjugate gradient descent methods. The latter approach minimizes the cost function $||Ax - y||$ through iterative updates to $x$. While gradient descent requires $O(1/\text{accuracy})$ iterations, sparse matrices often have high condition numbers, making the optimization sensitive and forcing a very small learning rate.

Let's consider performing a regular matrix-vector multiplication on the FPGA, using the "crankseg_2" matrix from Table 3 as an example. Assuming full utilization of the High Bandwidth Memory (HBM), each multiplication would take approximately 0.7s. This time is derived from the total number of float-32 elements ($63800^2 \times 2$) that need to enter the FPGA, and the FPGA bandwidth of 460GB/s. As for GUST, the preprocessing step takes 4.32s, and each subsequent SpMV (even if the vector changes) is completed in only 0.6ms.

### 5.4 Effect of Matrix Density and Structure on GUST

Figures 8(b)-(d) show that the speedup of GUST over 1D follows a $O(1/density)$ trend for each of the matrix structures. This suggests that hardware efficiency remains relatively

**Table 4. GUST vs Serpens** – Execution time (in terms of wall-clock time and clock cycles), energy consumption and throughput (GFLOPs/s) of GUST vs Serpens for the preprocessing and calculation phases on matrices introduced in Table 3. The preprocessing is done using `Intel core i7-10750H` CPU with a power consumption of 45 W.

| Matrix ID | Metric | GUST Pre. | GUST Calc. | Serpens Pre. | Serpens Calc. |
|---|---|---|---|---|---|
| (1) | Time | 4.32 s | 0.59 ms | 9.34 s | 0.93 ms |
|  | # Cycles | - | 57K | - | 208K |
|  | Energy | 194 J | 34.3 mJ | 420 J | 44.6 mJ |
|  | GFLOPS | - | 47.8 | - | 30.5 |
| (2) | Time | 4.56 s | 0.67 ms | 13.3 s | 0.85 ms |
|  | # Cycles | - | 64K | - | 190K |
|  | Energy | 205 J | 42.5 mJ | 598 J | 40.7 mJ |
|  | GFLOPS | - | 49.4 | - | 35.6 |
| (3) | Time | 4.72 s | 0.83 ms | 7.86 s | 0.73 ms |
|  | # Cycles | - | 80K | - | 163K |
|  | Energy | 212 J | 51.9 mJ | 353 J | 35.0 mJ |
|  | GFLOPS | - | 39.0 | - | 44.4 |
| (4) | Time | 8.69 s | 1.10 ms | 23.8 s | 1.37 ms |
|  | # Cycles | - | 106K | - | 306K |
|  | Energy | 391 J | 70.6 mJ | 1071 J | 65.7 mJ |
|  | GFLOPS | - | 50.1 | - | 42.2 |
| (5) | Time | 8.74 s | 1.45 ms | 26.2 s | 1.37 ms |
|  | # Cycles | - | 139K | - | 306K |
|  | Energy | 393 J | 91.1 mJ | 1179 J | 65.7 mJ |
|  | GFLOPS | - | 40.0 | - | 42.2 |
| (6) | Time | 9.79 s | 1.34 ms | 39.7 s | 2.09 ms |
|  | # Cycles | - | 129K | - | 466K |
|  | Energy | 440 J | 82.4 mJ | 1786 J | 100.4 mJ |
|  | GFLOPS | - | 31.5 | - | 31.1 |
| (7) | Time | 13.53 s | 1.52 ms | 70.0 s | 2.05 ms |
|  | # Cycles | - | 146K | - | 457K |
|  | Energy | 608 J | 91.6 mJ | 3150 J | 98.4 mJ |
|  | GFLOPS | - | 48.8 | - | 37.0 |
| (8) | Time | 5.58 s | 1.42 ms | 9.29 s | 1.87 ms |
|  | # Cycles | - | 136K | - | 417K |
|  | Energy | 251 J | 116.3 mJ | 417 J | 89.8 mJ |
|  | GFLOPS | - | 21.1 | - | 14.71 |
| (9) | Time | 10.97 s | 3.26 ms | 56.7 s | 4.52 ms |
|  | # Cycles | - | 313K | - | 1.01M |
|  | Energy | 493 J | 193.8 mJ | 2554 J | 217 mJ |
|  | GFLOPS | - | 19.8 | - | 14.29 |

stable across varying density levels, and that GUST's effectiveness is density-independent, which is in agreement with 12. However, the speedup varies between different matrix structures with the same density. GUST does a much better job dealing with matrices with uniform distribution than it does with power-law and K-mean. Depending on how well the NZ elements are spread out, we get a different standard deviation for #NZ elements in rows and column $mod\ l$ partitions (STD). As discussed in Section 3. 5, high STD negatively affects the performance of GUST. Load balancing helps reducing the high STD, but to some extent.

Energy efficiency gain also follows a $O(1/density)$ trend, but unlike speedup, it is not significantly affected by matrix structure. The reason for this is that NZ operations and data movement are the primary contributors to energy consumption, while the energy consumption from dynamic power during active device operation plays a less significant role.

### 5.5 Scalability

Looking at the energy efficiency gain of length-256 and -87 GUST with EC/LB in Figure 8, we notice that length-87 achieves a higher energy efficiency gain. This may seem surprising since length-256 offers more speedup. This is due to the fact that the power consumption of the crossbar connector in GUST increases superlinearly with length of GUST, as enumerated in Table 5. Furthermore, the hardware overhead introduced by the crossbar increases quadratically with length of GUST. This would be problematic if we were to use a long length GUST. One possible solution to address this problem is to use a parallel arrangement of multiple GUSTs rather than one big GUST; i.e., use $k$ length-$l$ GUSTs in parallel rather than one length-$kl$ GUST. GUST naturally gets parallelized really well since it performs by taking in a set of rows, calculating the corresponding vector elements, and then taking the next set of rows. Each set of rows will be processed indepenetly and in parallel. Moreover, the Edge-Coloring schedule would not need to change (schedule found for length-$l$ GUST is applicable to $k$ parallel length-$l$ GUSTs).

Although the parallel arrangement reduces the hardware overhead and energy consumption, it will increase the execution for two reasons: 1. The main purpose of GUST's hardware is to enable resource sharing among multiple rows and columns. If we replace a length-$kl$ GUST with length-$l$, we virtually decrease resource sharing among $kl$ rows and $kl$ columns to resource sharing among $l$ rows and $l$ columns. 2. It is not guaranteed that the work will be divided equally among the parallel GUSTs. Other factors such as number of arithmetic units and BW remains the same between the parallel and single arrangement of GUST.

### 6 Related Work

Many recent studies have shifted their focus towards accelerating SpMV of FPGAs. For example, Copernicus [2] investigates the effects of employing various compression formats, including LIL, for SpMV implementation on FPGAs. ReDESK [21] proposes a representation specifically designed for data prefetching on CPUs, allowing streaming processing on FPGAs. Recent studies have also examined the computation order, such as a work by Li et al.[19], which reorganizes the nonzero elements to enhance data reuse, thereby optimizing memory requests. Customizing network architecture using elementary blocks for low-level detail exploitation has been explored in an FPGA work [13], along with the use of tree structures like in Two-Step [27] and Fafnir [1] for memory streaming. Sparstition [28] looks into the acceleration of SpMV by examining partitioning schemes. Tree structures have been utilized by several SpMV accelerators, such as MeNDA [9], a scalable multi-way merge accelerator near DRAM that enhances performance and reduces energy consumption. It adopts a merge sort-based algorithm with a high-performance hardware merge tree and various techniques to exploit increased system memory bandwidth, achieving remarkable speedup and efficiency gains. A different tree-based method includes an accelerator design focusing on high-bandwidth utilization [20], with elaborate FPGA implementation showing promising performance improvements.

**Table 5. Per-partition Resource Consumption of GUST–**
Resource consumption of length-8, -87 and -256 GUST in terms of power (W) and number of units used by the partitions of GUST. The arithmetic and I/O partitions of GUST have a resource consumption increasing linearly with GUST's length while the crossbar connector has a quadratic and superlinearly scaling hardware and power consumption.

| Seg- | length 8 | | length 87 | | length 256 | |
|---|---|---|---|---|---|---|
| ment | Power | Units | Power | Units | Power | Units |
| Arith-metic | 0.3 | LUT: 4229<br>Reg: 256<br>DSP: 16<br>Carry8: 152 | 3.5 | LUT: 46.0K<br>Reg: 2.8K<br>DSP: 174<br>Carry8: 1.6K | 6.3 | LUT: 132K<br>Reg: 8.2K<br>DSP: 512<br>Carry8: 4.8K |
| Cross-bar | 1.0 | LUT: 772<br>Reg: 256 | 3.6 | LUT: 17.3K<br>Reg: 2.8K | 16.4 | LUT: 756K<br>Reg: 8.2K |
| IO | 0.5 | IO Pins: 802<br>Buff: 546 | 7.1 | IO Pins: 8.9K<br>Buff: 6.2K | 28.1 | IO Pins: 27K<br>Buff: 18K |

Additionally, a hardware/algorithm co-optimized accelerator for extensive SpMV problems is presented in a recent paper [26], which proposes a data transfer-efficient algorithm coupled with a specialized hardware model that includes an ASIC for multi-way merge operation and integrates state-of-the-art 3D stacked HBM. SpaceA [35], an SpMV accelerator, leverages processing-in-memory (PIM) architectures and integrates features to overcome irregular memory access patterns, resulting in substantial speedup and energy savings. A novel streaming implementation for coordinate format (COO) sparse matrix-vector multiplication shows up to 6× speedup and higher energy efficiency in another work [24]. A study identifying challenges in developing high-performance sparse linear algebra accelerators on HBM-equipped FPGAs is presented in [8]. The case study on SpMV leads to HiSparse, an accelerator displaying promising speedup and bandwidth efficiency over existing solutions. Serpens [29], another HBM-based FPGA accelerator, is designed for general-purpose sparse matrix-vector multiplication, featuring memory-centric processing engines.

## 7 Conclusions and Future Work

Through innovative hardware/software co-design, GUST achieves a hardware utilization of 33.67%, and speedup and energy efficiency gain of 411× and 137× over 1D systolic array using the same number of arithmetic units for SpMV. Compared to Serpens, a state-of-the-art FPGA-based SpMV accelerator, GUST achieves lower execution time for seven out of nine matrices, and lower energy consumption for five of the matrices. This performance is achieved through resource sharing among multiple rows and columns, which is enabled through separating partial sum and partial product calculation in the hardware, and eliminating collisions with edge-coloring scheduling in the software. Furthermore, the effectiveness is independent from sparsity sparsity, and can be utilized in a parallel manner, making GUST a versatile accelerator that can be used for a wide range of applications. One possible direction for GUST would be to investigate the possibility of resource sharing in a 2-dimensional GUST for sparse matrix-matrix multiplication. Moreover, the notion of edge-coloring could be extended to other applications with shared resources for removing collisions.

The ideas presented here is not limited to FPGAs only, and is applicable to any hardware platform that can provide a set of multipliers and adders, and a crossbar connector. For example, consider GPUs. Each block of threads in a GPU has a shared memory that functions as a crossbar connector by design. Each thread calculates a partial product and stores it in the shared memory. The threads then synchronize and access the shared memory to update the partial sums. Since a GPU is structured as multiple blocks of threads, each with a limited amount of shared memory, the implementable GUST is a small length-k GUST for each block. Of course there are more details to consider, especially since GPUs are often memory-bound in the case of matrix-vector multiplication.